# Ether: Bitcoin's competitor or ally?

Jamal Bouoiyour [1] and Refk Selmi [2]

**Abstract:** Although Bitcoin has long been dominant in the crypto scene, it is certainly not alone. Ether is another cryptocurrency related project that has attracted an intensive attention because of its additional features. This study seeks to test whether these cryptocurrencies differ in terms of their volatile and speculative behaviors, hedge, safe haven and risk diversification properties. Using different econometric techniques, we show that a) Bitcoin and Ether are volatile and relatively more responsive to bad news, but the volatility of Ether is more persistent than that of Bitcoin; b) for both cryptocurrencies, the exuberance and the collapse of bubbles were identified, but Bitcoin appears more speculative than Ether; c) there is negative and significant correlation between Bitcoin/Ether and other assets (S&P500 stocks, US bonds, oil), which would indicate that digital currencies can hedge against the price movements of these assets; d) there is negative tail independence between Bitcoin/Ether and other financial assets, implying that these cryptocurrencies exhibit the function of a weak safe haven; and e) The inclusion of Bitcoin/ Ether in a portfolio improve its efficiency in terms of higher reward-to-risk ratios. But investors who hold diversified portfolios made of stocks or bonds and Ether may face losses over bearish regime. In such situation, stock and bond investors may take a short position on Bitcoin

---

[1] CATT, University of Pau, Avenue du Doyen Poplawski, 64016 Pau Cedex, France. E-mail : jamal.bouoiyour@univ-pau.fr
[2] University of Tunis, Campus Universitaire, Avenue 7 Novembre, 2092, Tunis, Tunisia; University of Pau, France. Email: s.refk@yahoo.fr

# 1. Introduction

Forty eight years ago the first data packets were sent across the network that became what we know as the internet. At this date, everyone was talking about the power of internet, and its potential impacts on our lives, but no one imagine that our lives will change fundamentally. Nowadays, some expect that the Blockchain has the power to reinvent key institutions. Ten years from now we will wonder how institutions and businesses could have been survived without the internet of value.

The blockchain rose in the onset of the global financial collapse, when an anonymous programmer under the pseudonym Satoshi Nakamoto released a new protocol for "A Peer-to-Peer Electronic Cash System" using a cryptocurrency, namely Bitcoin. In the wake of the global financial crisis, policymakers faced substantial challenges as the financial markets were in turmoil, credits flows were disrupted and the economies moved into deep recession. In response, some central banks – in particular, the U.S. Federal Reserve Bank (Fed), the Bank of England (BoE), and the European Central Bank (ECB) – have embarked on ever-more expansionary monetary policies while trying to avoid falling into deflation. Central banks planners asserted banks needed bailouts to mitigate the risk of deflationary spirals. When all "normal" tools of monetary policy were used and seemed unsuccessful to drive down long-term interest rates and spur their economies, the pressure to use more "aggressive" monetary instruments raises; hence the usefulness of something known as "quantitative easing" (QE). This instrument aims at putting downward pressure on real long-run interest rates, bolstering prices for corporate equities, enhancing aggregate demand, mitigating disinflationary pressures, and stimulating overall financial conditions (Engen et al. 2015). It must be stressed that central banks ordinarily pursue monetary policy by buying and selling short-term debt securities to target short-term nominal interest rates. These purchases and sales of assets significantly affect the monetary base. In other words, there a two ways thereby a central bank can expand the monetary base by buying bonds from the public, or by lending money to the public. By increasing the monetary base, central banks can affect a variety of asset prices, including exchange rates and stock prices. Making more money available is assumed to encourage financial institutions to lend more to businesses, pushing down the interest rates. Favorable financial conditions would, in turn, help to improve aggregate demand and avoid disinflationary pressures by reinforcing support for consumer spending and enhancing investment environment. But the money creation has not yet found out its way back to the ordinary citizens, and the stimulus packages that were anticipated to ease better liquidity into global markets do not occur systematically. Overall, both conventional (i.e., manipulation of interest rates) and unconventional policies (i.e., quantitative easing) have demonstrated their inefficiency to stimulate economic growth in a deflationary and uncertain context. These considerations have led to a trend towards questioning the effectiveness of standard economic and financial structures which govern the conventional monetary and financial system. This has lessened the fiat's ability to continue to hold value. Here, the digital currencies (in particular, Bitcoin) are leading the charge by providing a completely decentralized secure alternative to fiat currencies during times of economic and geopolitical chaos, representing therefore a possible remedy to protect individuals from inflation pressures and devaluation.

Cryptocurrencies are dissimilar from traditional fiat currencies since no government issues or controls them (Bucholz et al. 2012 and Yermack 2014, Bouoiyour et al. 2016, among others). Bitcoin is virtual money with zero intrinsic value issued by computer code in electronic portfolios, which is not convertible into anything and not have the backing of any Central Banks and any government. The value of a Bitcoin cannot be considered as convertible tangible asset (such as gold) or a fiat currency (such as dollar). It is determined by the interplay of supply and demand. Since the creation of Bitcoin in 2009, the idea of exploiting its enabling technology to develop applications beyond currency has been achieving a wide level of international recognition. It facilitates business transactions from person to person worldwide without any intermediary, reduces trade barriers and increases the productivity. But it remains far from certain for many reasons including its extra-volatility, its speculative behavior, its inelastic money supply coded by mathematic formula and the lack of legal security (Bouoiyour and Selmi 2015; Ciaian et al. 2016). In addition to Bitcoin, a number of alternative platforms have flourished over the last few years. Due to its status as the original blockchain protocol, it should be not surprising that Bitcoin has long dominated the cryptocurrency markets. However, there's also sharp evidence that this long-standing record could be changing. Currently, the world's leading cryptocurrency (i.e., Bitcoin) faces growing competition from Ethereum. The latter is an open-source and public blockchain that anyone can employ as a decentralized ledger. It was developed with the main purpose of creating a more generalized blockchain platform, enabling to straightforwardly build applications that benefit from the decentralization and security properties of blockchains, and to evade the necessity to generate a novel blockchain for each new application. The ethereum blockchain has its own cryptocurrency called Ether , which is similar to Bitcoin, but what attracts the attention of several companies is the underlying Ethereum network. Even though the Bitcoin blockchain has tended to be utilized for payment transactions, the adoption of Ethereum blockchain technology by the corporate world implies it could be much larger than its early stage rival. Ethereum technology is expected to highly enhance smart contract applications that can make automatic intricate physical and financial supply chain procedures.

While everyone in 2017 was focused on the astonishing growth of Bitcoin, another cryptocurrency namely Ether has been quietly rising. The value of Ether has increased by about 4,500 percent since the start of the year. The industry publication CoinDesk claimed in June 2017 and based on a survey of 1,100 virtual currency users, that 94 percent were optimistic about the situation of Ethereum and its related cryptocurrency (Ether), while only 49 percent were positive about the state of Bitcoin. Ethereum uses the same technology as Bitcoin does, i.e. the Blockchain. Nevertheless, Ethereum does not settle for just being a virtual currency (Ether). It employs the Blockchain to certify contracts in the quickest and safest way possible (i.e., smart contracts). It is an ecosystem allowing the creation of decentralized applications. With Ethereum, every people are the owner of his personal data with free information while limiting the possibility of frauds. In this ground, we try to test if Ethereum's cryptocurrency (i.e., Ether) and Bitcoin are competitors. More specifically, we will address whether Bitcoin's market dominance is being challenged, and if Ether is the digital currency of the moment. The replies to these questions depend on how you will compare them. Although it is clearer that Bitcoin and Ethereum blockchains differ in terms of purpose, supply, security and mainstream adoption (see Table A.1, Appendix), it is still unclear whether Bitcoin and Ether differ regarding their volatilities, speculative behaviors, risk diversification, hedge and safe haven properties. Using different econometric techniques (Optimal GARCH model, the generalized sup ADF test procedure, different bivariate copulas and Reboredo (2013)'s risk diversification methodology), we show that Bitcoin is less volatile and more speculative than Ether. This analysis highlights also the usefulness of both Bitcoin and Ether as hedge, weak safe haven and portfolio diversifier.

The outline of the paper is as follows: Section 2 describes the methodology and data. Section 3 reports and discusses the results. Section 4 concludes.

## 2. Methodology and data

This paper aims at comparing Bitcoin and Ether across five features: volatility, speculation, hedge, safe haven and portfolio diversification capabilities. For this purpose, we first use many GARCH extensions to adequately estimate the volatility of Bitcoin and ether prices. Second, we pay special attention to the explosive nature of bubbles by using a new econometric method, namely the generalized sup ADF test procedure developed by Phillips et al. (2013). Third, we explore the dependence structure between Bitcoin/Ether prices and and other financial assets using various copula functions with symmetric and asymmetric tail behaviors. To address whether cryptocurrencies serve as hedge or safe haven investments, we follow the existing literature on the properties of precious metals (in particular, gold). To test if gold exhibits hedge and safe haven properties safe haven, several studies have examined the connection between gold and stocks, bonds and oil. For example, Coudert and Raymond (2010) explored the role of gold as a hedge and safe haven against US stocks during recessions and bear markets. Guimaraes (2013) focused on the specific role of gold as a safe haven against US bonds. Robodero (2013) investigated the role of gold as a hedge or safe haven against oil price movements, and found that gold cannot hedge against oil price movements, but it can act as an effective safe haven against extreme oil price movements. Another reason behind the choice of US stocks and bond as explanatory variables is that the co-movements among assets is substantially due to innovations in the global factors, which U.S. fundamentals may have a wider role to play. To this end and after comparing Bitcoin and Ethereum in terms of their volatilities and the speculative behaviors, we assess the dependence between Bitcoin/Ether prices (*BPI* and *ETH*, respectively) and S&P500 stocks (*STR*) and bonds (*BdR*) and the real crude oil price (*Oil*). Fourth, we analyze the potential reduction in the risk portfolio generated by the inclusion of digital currencies (Bitcoin vs. Ether) in portfolios composed by stocks, bonds and oil price.

### 2.1. GARCH-type modeling

Volatility clustering and leptokurtosis are commonly observed in economic and financial time series. Other phenomena usually encountered are the so- called "leverage effect" and "nonlinear effect". The GARCH (General Autoregressive Conditional Heteroskedasticity)-type modeling has been and continuous to be very valuable tool in finance and economics since the seminal paper of Engle (1982). Engle (1982) proposed to model time-varying conditional variance with Auto- Regressive Conditional Heteroskedasticity (ARCH) processes using lagged disturbances. He argued that a high ARCH order is required to properly capture the dynamic behaviour of conditional variance. The Generalized ARCH (GARCH) model of Bollerslev (1986) fulfills this requirement as it is based on an infinite ARCH specification which minimizes the number of estimated parameters, denoted as:

$$\sigma_t^2 = \omega + \sum_{i=1}^{q} \alpha_i \varepsilon_{t-i}^2 + \sum_{i=1}^{p} \beta_j \sigma_{t-j}^2 \qquad (1)$$

Where $\alpha_i$, $\beta_i$ and $\omega$ are the parameters to estimate.

Although the ARCH and GARCH models detect volatility clustering and leptokurtosis, their distributions are symmetric and linear. In other words, they do not account for possible asymmetry and nonlinearity in the volatility dynamics. To address these problems, we apply several GARCH extensions, such as the Exponential GARCH (EGARCH) model by Nelson (1991), the Asymmetric Power ARCH (APARCH) model by Ding et al (1993), the weighted GARCH model by Bauwens and Storti (2008), among others. Table A.2. (Appendix) reviews succinctly the different GARCH models used throughout this study. Since no single measure of volatility has dominated the existing empirical literature, the appropriate model able to properly depict the volatile behavior Bitcoin and ether prices can be selected using standard criteria such as the Akaike information criterion (AIC), the Bayesian (BIC) and Hannan-Quinn information criteria (HQ). These criteria are sufficient to judge the quality of conditional variance estimation in terms in terms of trade-off between goodness of fit and model parsimony.

### 2.2. Detection of bubble periods

To identify periods of bubble, we use a new econometric tool developed by Phillips and Yu (2011) and Phillips et al. (2011), and then extended in a generalized form of the sup Augmented Dickey Fuller (GSADF) by Phillips et al. (2013). The main consideration in defining explosive periods are controlling for structural breaks that may yield to the non-rejection of the unit-root hypothesis (Perron 1989). To resolve this problem, Gil-Alana (2003) assumed well known structural break dates in their examination, whereas Gil-Alana (2008) applied a residuals sum squared approach where a single structural break date is accounted for at an unknown time. This study recursively determines, via a flexible moving sample test procedure (GSADF test), periods where the lower bound of the fractional order exceeds unity (bubble periods), and subsequently return to levels below unity (stable periods), enabling us to adequately capture and date-stamp explosive periods. In brief, this approach considers multiple structural breaks at unknown dates (Balcilar et al. 2015). Based on this method, bubbles are detected in a consistent manner even with smaller sample sizes (Phillips et al. 2013; Caspi et al. 2015).

The Phillips et al. (2013)'s test procedure performed throughout this research recursively implements an ADF-type regression test through a rolling window procedure. Suppose the rolling interval starts with a fraction $r_1$ and ends with a fraction $r_2$ of the total number of observations, with the size of the window $r_w=r_2-r_1$, then let:

$$y_t = \mu + \delta y_{t-1} + \sum_{i=1}^{p} \phi^i_{r_w} \delta y_{t-1} + \varepsilon_t \tag{2}$$

where $\mu$, $\delta$ and $\phi$ are the parameters to be estimated via OLS regression, and the usual $H_0$: $\delta = 1$ then tested against the right sided alternative $H_1$: $\delta > 1$. The number of observations under consideration is $T_w = [r_w, T]$, where [.] is the integer part. The ADF statistic corresponding to 1 is expressed by $ADF_{r_1}^{r_2}$.

Phillips et al. (2013) proposed a backward sup ADF test where the end point of the subsample is fixed at a fraction $r_2$ of the whole sample and the window size is extended from the fraction $r_0$ to the fraction $r_2$. Thus, the backward sup ADF statistic is denoted as:

$$SADF_{r_2}(r_0) = \sup_{r_1 \in [0, r_2 - r_1]} ADF_{r_1}^{r_2} \tag{3}$$

The key reason behind using a sup ADF statistic is the fact that CDS price bubbles may collapse temporarily, and thus the standard unit root tests may have a restricted power in capturing bubble-periods (Caspi et al. 2015). In this context, Homm and Breitung (2012) claimed that the sup ADF test procedure seems suitable in bubble-detection purpose, especially when dealing with one or two bubble episodes. The GSADF is constructed by re-testing the SADF test procedure for each $r_2 \in [r_0, 1]$.

$$GSADF(r_0) = \sup_{r_2 \in [r_0, 1]} SADF_{r_{21}}(r_0) \tag{4}$$

The GSADF corresponds to a sequence of ADF statistics. The supremum value of this sequence (SADF) is utilized to test the null hypotheses of unit root against its right-tailed (mildly explosive) alternative while comparing it to its corresponding critical values. Generally speaking, the testing procedure discussed above is pursued to test whether UK and European CDS spreads exhibit bubble patterns within a specific sample. When we note significant ADF statistics (i.e. $\delta_{r_1, r_2} \succ 1$), we can deduce that there exist bubble periods. If the null hypothesis of no bubbles is rejected, the Phillips et al. (2013)'s test allows to date-stamping the beginning and the ending points of the explosive episodes. The starting point of a bubble corresponds to the date, expressed as $T_{r_e}$ at which the backward sup ADF sequence crosses the critical value from below. Likewise, the ending point of a bubble is also defined as the date, written as $T_{r_f}$ at which the backward sup ADF sequence crosses the critical value but from above. Ultimately, based on GSADF, the explosive periods can be denoted as:

$$\hat{r}_e = \inf_{r_2 \in [r_0, 1]} \left\{ r_2 : BSADF_{r_2} \succ cv_{r_2}^{\beta_T} \right\} \quad (5)$$

$$\hat{r}_f = \inf_{r_2 \in [\hat{r}_e, 1]} \left\{ r_2 : BSADF_{r_2} \succ cv_{r_2}^{\beta_T} \right\} \quad (6)$$

where $cv_{r_2}^{\beta_T}$ is the $100(1-\beta_t)\%$ critical value of the sup ADF statistic based on $[T_{r_2}]$ observations. We set $\beta_t$ to a constant value, 5%, as opposed to letting $\beta_t \to 0$ as T → 0. Note that the $BSADF$ ($r_0$) for $r_2 \in [r_0, 1]$ is the backward sup ADF statistic that relates to the GSADF statistic, and denoted as: $GSADF(r_0) = \sup_{r_2 \in [r_0, 1]} \{BSADF_{r_2}(r_0)\}$

### 2.3. Copula models

The Bitcoin and Ether climbs alongside the great uncertainty surrounding the World in 2017 (the Brexit, the Trump's win in US presidential elections, China's deepening slowdown, India demonetization, demonetization in Venezuela, the elections in Europe including France, Germany and the Netherlands, see Bouoiyour and Selmi 2017) has led some to proclaim them as "digital gold" or new hedges/safe havens. We attempt here to test this hypothesis by examining the dependence structure between Bitcoin/Ether and other assets. To do so, we consider various copula functions with symmetric and asymmetric tail behavior, even if we control for time-varying dependence. First, we consider elliptical Gaussian and Student-t copulas which are usual choices for the market dependence structure. These functions are denoted, respectively, as:

$$C^{Gaussian}(u_t, v_t; \rho) = \Phi(\Phi^{-1}(u_t), \Phi^{-1}(v_t)) \quad (7)$$

$$C^{Student-t}(u_t, v_t; r, v) = T_v(t_v^{-1}(u_t), t_v^{-1}(v_t)) \quad (8)$$

where $\Phi$ is the bivariate standard normal CDF with correlation $\rho$ ($-1 < \rho < 1$); $\Phi^{-1}(u_t)$ and $\Phi^{-1}(v_t)$ are standard normal quantile functions; $T$ is the bivariate Student-t CDF with degree-of-freedom parameter $v$ and correlation $\rho$ ($-1 < \rho < 1$); and $t_v^{-1}(u_t)$ and $t_v^{-1}(v_t)$ are the quantile functions of the univariate Student-t distributions. Both copulas display symmetric dependence, even though the Gaussian has zero tail dependence and the Student-t displays tail dependence given by $\lambda_U = \lambda_L = 2t_{v+1}(-\sqrt{v+1}\sqrt{1-\rho}/\sqrt{1+\rho}) > 0$.

We consider copulas with symmetric tail dependence, namely, the Plackett and the Frank copulas, expressed, respectively, as:

$$C^{Plackett}(u_t, v_t; \pi) = \frac{1}{2(\pi-1)}(1+(\pi-1)(u_t+v_t) - \sqrt{(1+(\pi-1)(u_t+v_t))^2 - 4\pi(\pi-1)u_t v_t}) \quad (9)$$

$$C^{Frank}(u_t, v_t; \lambda) = \frac{-1}{\lambda}\log\left(\frac{(1-e^{-\lambda}) - (1-e^{-\lambda u_t})(1-e^{-\lambda v_t})}{(1-e^{-\lambda})}\right) \quad (10)$$

where $\pi \in [0, \infty) \setminus \{1\}$ and $\lambda \in (-\infty, \infty) \setminus \{0\}$. It must be mentioned that the two above copulas exhibit tail independence.

Due to the fact that the relationship between the variables of interest may behave dissimilarly over different market circumstances (for example, booms versus bursts), this study performs copula functions with asymmetric tail dependence structures. The Gumbel copula reflects upper tail dependence, while its rotation reflects lower tail dependence, expressed, respectively, as:

$$C^{Gumbel}(u_t, v_t; \delta) = \exp\left(-\left((-\log u_t)^\delta + (-\log v_t)^\delta\right)^{1/\delta}\right) \quad (11)$$

$$C^{Rotated\_Gumbel}(u_t, v_t; \delta) = u_t + v_t - 1 + C^{Gumbel}(1-u_t, 1-v_t; \delta) \quad (12)$$

where $\delta \in (1, \infty)$. The upper and lower tail dependence structures of the Gumbel copula are $\lambda_U = 2 - 2^{1/\delta}$ and $\lambda_L = 0$, respectively, while the opposite holds for the rotated Gumbel. We also consider the symmetrized Joe-Clayton (SJC) copula as it takes into account both lower and upper tail dependence, indicating the occurrence or not of asymmetry. It is expressed as follows:

$$C^{SJC}(u_t, v_t; \lambda_U^{SJC}, \lambda_L^{SJC}) = 0.5(C^{JC}(u_t, v_t; \lambda_U^{JC}, \lambda_L^{JC}) + C^{JC}(1-u_t, 1-v_t; \lambda_U^{JC}, \lambda_L^{JC}) + u_t + v_t - 1) \quad (13)$$

where $C^{JC}(u_t, v_t; \lambda_U^{JC}, \lambda_L^{JC}) = 1 - (1 - \{[1-(1-u_t)^\kappa]^{-\gamma} + [1-(1-v_t)^\kappa]^{-\gamma} - 1\}^{-1/\gamma})^{1/\kappa}$, $\kappa = 1/\log_2(2 - \lambda_U^{JC})$ and $\gamma = -1/\log_2(\lambda_L^{JC})$. Moreover, $\lambda_U^{SJC}(v) \in (0,1)$ and $\lambda_L^{SJC}(v) \in (0,1)$. For this copula function, the tail dependence coefficients are themselves the parameters of the copula. If $\lambda_U^{SJC} = \lambda_L^{SJC}$, then the market structure is symmetric, otherwise it is asymmetric.

We then account for possible time-varying dependence structure. For the Gaussian and Student-t copulas, we depict the dynamics of the linear dependence parameter as evolving flexibly over time based on a new econometric tool proposed by Patton (2006), written as follows:

$$\rho_t = \Lambda\left(\Psi_0 + \Psi_1 \rho_{t-1} + \Psi_2 \frac{1}{10} \sum_{j=1}^{10} \Phi^{-1}(u_{t-j}) \cdot \Phi^{-1}(v_{t-j})\right) \tag{14}$$

where $\Lambda$ denotes the logistic transformation $\Lambda(x) = (1-e^{-x})(1+e^{-x})^{-1}$ that is employed to keep $\rho_t$ within (-1,1). For the Student-t copula, $\Phi^{-1}(x)$ is substituted by $t^{-1}_v(x)$. For the conditional Gumbel copula and its rotation, the evolution of $\delta$ is specified based on ARMA(1,10) process, that can be expressed as following:

$$d_t = \Psi_0 + \Psi_1 d_{t-1} + \Psi_2 \frac{1}{10} \sum_{j=1}^{10} |u_{t-j} - v_{t-j}|. \tag{15}$$

For the SJC copula, the variation of upper and lower tail dependencies over time can be given by an ARMA(1,10) process denoted, respectively, as:

$$\tau_{U,t}^{SJC} = \Lambda\left(\Psi_0^U + \Psi_1^U \tau_{U,t-1}^{SJC} + \Psi_2^U \frac{1}{10} \sum_{j=1}^{10} |u_{t-j} - v_{t-j}|\right) \tag{16}$$

$$\tau_{L,t}^{SJC} = \Lambda\left(\Psi_0^L + \Psi_1^L \tau_{L,t-1}^{SJC} + \Psi_2^L \frac{1}{10} \sum_{j=1}^{10} |u_{t-j} - v_{t-j}|\right) \tag{17}$$

### 2.4. Portfolio risk management

The diversification enables investors to obtain a desired return without taking as much risk as with an individual security. Thus, a good diversifier is a portfolio addition that helps to minimize the overall risk in a portfolio. This study tries to answer which of digital currencies (Bitcoin or Ether) is believed to have this quality? The exercise consists of analyzing the potential reduction in the portfolio risk generated by the inclusion of Bitcoin and Ether in portfolios composed by some assets (in particular, S&P500 stocks, US bonds and oil price). Following Reboredo (2013), we evaluate the usefulness of virtual currencies for portfolio risk management by comparing the risk of a benchmark portfolio (Portfolio 1) composed only by S&P500 stocks with the risks for portfolios composed of Bitcoin and S&P500 shares (Portfolio 2) and another that is formed by Ether and S&P500 stocks (Portfolio 3). We follow the same procedure by comparing the risk portfolio that includes only US bonds (portfolio 4) with the portfolios containing Bitcoin and US bonds (portfolio 5) and Ether and US bonds (portfolio 6). Ultimately, we compare a benchmark portfolio (Portfolio 7) containing Oil with portfolios incorporating Bitcoin and Oil (portfolio 8) and Ether and Oil (portfolio 9). We consider, respectively, risks-minimizing *BPI-STR, ETH-STR, BPI-BdR and ETH-BdR, BPI-Oil* and *ETH-Oil* portfolios. According to Kroner and Ng (1998)'s study, at time *t*, the optimal weights of Bitcoin/Ether prices in these different portfolios ($w_t^{BPI}$ and $w_t^{ETH}$) are written as follows:

$$w_t^{BPI} = \frac{h_t^{STR} - h_t^{BPI,STR}}{h_t^{BPI} - 2h_t^{BPI,STR} + h_t^{STR}} \; ; w_t^{ETH} = \frac{h_t^{STR} - h_t^{ETH,STR}}{h_t^{ETH} - 2h_t^{ETH,STR} + h_t^{STR}} \quad (18)$$

$$w_t^{BPI} = \frac{h_t^{BdR} - h_t^{BPI,BdR}}{h_t^{BPI} - 2h_t^{BPI,BdR} + h_t^{BdR}} \; ; w_t^{ETH} = \frac{h_t^{BdR} - h_t^{ETH,BdR}}{h_t^{ETH} - 2h_t^{ETH,BdR} + h_t^{BdR}} \quad (19)$$

$$w_t^{BPI} = \frac{h_t^{Oil} - h_t^{BPI,Oil}}{h_t^{BPI} - 2h_t^{BPI,Oil} + h_t^{Oil}} \; ; w_t^{ETH} = \frac{h_t^{Oil} - h_t^{ETH,Oil}}{h_t^{ETH} - 2h_t^{ETH,Oil} + h_t^{Oil}} \quad (20)$$

where $h_t^i$ are the conditional volatility of *i* (where *i* is respectively Bitcoin price, Ethereum price, S&P500 stocks, US bonds and oil price), $h_t^{BPI,STR}$, $h_t^{ETH,STR}$, $h_t^{BPI,BdR}$, $h_t^{ETH, BdR}$ and $h_t^{BPI,Oil}$ and $h_t^{ETH,Oil}$ are respectivelyof, the conditional covariance between the Bitcoin price and S&P500 stocks, the Ethereum price and S&P500 stocks, the Bitcoin price and US bonds, the Ethereum price and US bonds, the oil price and S&P500 shares and the Ethereum price and US bonds . For each pair, the information needed relies on computing $w_t^{BPI}$ and $w_t^{ETH}$ derived from copula models discussed above.

Thereafter, the risk reduction effectiveness is assessed by comparing the percentage reduction in the variances of portfolios 2 and 3 against the Portfolio 1, the variances of portfolios and 5 and 6 with portfolio 4, and the variances of portfolios and 8 and 9 with portfolio 7 as demonstrated in the following:

$$RE_{Var} = 1 - \frac{Var(P_j)}{Var(P_i)} \quad (21)$$

where *i* corresponds to portfolios 1, 4 and 7; and *j* corresponds to portfolios 2, 3, 5, 6, 8 and 9 and $Var(P_j)$ and $Var(P_i)$ are the variances of the portfolios *j* and portfolios *i*, respectively; $RE_{Var}$ takes values fluctuating between 0 and 1, where a higher value implies a strong variance reduction.

## 2.5. Data

The financial data set used in our empirical estimations, consists of daily data, from 01 August 2015 to 30 June 2017 (700 observations)[3], on Bitcoin price, Ether price, US stocks and bonds and real oil price. The Bitcoin price index (*BPI*) is an index of the exchange rate between the US dollar (USD) and the Bitcoin. The CoinDesk Bitcoin Price Index represents an average of Bitcoin prices across leading bitcoin exchanges. The Ether price index (*ETH*) is an index of the exchange rate among USD and the Ether. US stock and bond log returns are calculated from the S&P500[4] and bond indices with returns being computed as the first-differences of the natural logs of these indices (*STR* and *BdR*, respectively). Table 1 reports all the data used and the sources.

**Table 1. Data sources**

|  | Variables | Definition | Sources |
|---|---|---|---|
| The dependent variables | *BPI* | Bitcoin price index | CoinDesk (www.coindesk.com/price) |
|  | *ETH* | Ether price index |  |
| The determinants | *STR* | S&P500 price returns | Global Financial Database |
|  | *BdR* | US Bond price returns | (https://www.globalfinancialdata.com/ ) |
|  | *Oil* | The real crude oil price changes | U.S. Department of Energy (https://energy.gov/) |

We transformed all the variables by taking natural logarithms to correct for heteroskedasticity and dimensional differences. Descriptive statistics for series are reported in Table 2. We note that Ether has the most sizeable volatility, followed by Bitcoin. The returns are non-normal as indicated by the Jarque-Bera tes. Also notably, both Ether and Bitcoin exhibits a positive skewness. What does that imply? Skewness is an asymmetry from the normal distribution. In other words, one side of distribution does not mirror the other side5. Even though our results indicate that US stock and bond returns and oil price changes are negatively skewed, Bitcoin and Ether are exceptionally positively skewed, which means that incorporating Bitcoin and Ether in a portfolio may improve the portfolio's skewness (i.e. mitigate harmful risks). It means that in the Bitcoin and Ether markets, there is a large chance of a 10% increase in one day than there is a 10% fall any other day. In brief, these findings underscore the roles that may play Bitcoin and Ether as good hedge in periods of distress.

---
[3] The choice of this period is due to the availability of data, in particular the Ether price index data.
[4] The S&P 500 is largely viewed as the best single gauge of large-cap U.S. shares. This index is composed by 500 leading companies and captures approximately 80% coverage of available market capitalization.

**Table 2. Some statistical properties of variables under study**

|  | *BPI* | *ETH* | *S&P500* | *US bonds* | *Oil* |
|---|---|---|---|---|---|
| Mean | -0.0010 | -0.0002 | 0.0006 | -0.0001 | -0.0011 |
| Std. Dev. | 0.1026 | 0.1188 | 0.0721 | 0.0878 | 0.0817 |
| Skewness | 1.3358 | 1.4971 | -1.2229 | -3.4612 | -0.5504 |
| Kurtosis | 27.927 | 5.6098 | 11.555 | 13.452 | 9.2572 |
| Jarque Bera | 513.42 | 647.95 | 505.07 | 438.29 | 711.23 |

## 3. Results
### 3.1. Volatility

To choose the "optimal" GARCH model able to measure the volatility of Bitcoin and Ether prices, we use standard historical evaluation criteria (Akaike, Bayesian and Hannan-Quinn criteria). Whatever the criterion employed, the GARCH extension chosen to depict the volatility of Bitcoin price is the Exponential-GARCH (see Table A.3, Appendix). One of the most important limitations of standard GARCH models is that they are unable to capture the stylized fact that conditional variance tends to be more pronounced after a decrease in return than after an increase. So, to control for asymmetry, many alternative models have been proposed including the Exponential- GARCH (E-GARCH) introduced by Nelson (1991). This model specified the conditional variance in logarithmic form denoted as:

$$\log(\sigma_t^2) = \omega + \sum_{i=1}^{q}(\alpha_i z_{t-i} + \gamma_i(|z_{t-i}| - \sqrt{2/\pi})) + \sum_{i=1}^{p}\beta_j \log(\sigma_{t-j}^2) \tag{22}$$

where $\alpha_i$, $\beta_j$, $\omega$, $\gamma$ are the parameters to estimate, and $z_t$ the standardized value of error.

---

5 There is negative and positive skewness conditioning upon whether data points are skewed to the left (negative skew) or to the right (positive skew) of the data average.

The estimates are reported in Table31. We find that the leverage effect is positive and significant, indicating that the volatility of Bitcoin price is typically more responsive to bad news. The duration of volatility's persistence appears stronger ($\alpha + \beta + 0.5\gamma = 0.87$).

For Ether price, the best GARCH model chosen based on the same information criteria is the Threshold-GARCH model (T-GARCH, Table A.3, Appendix). In most widely used GARCH models the conditional variance is defined as a linear function of lagged conditional variances and squared past returns. Though these models have been proved to be adequate for capturing the dependence structure in conditional variances, they contain important limitations, one of which is that they fail to detect structural breaks that may stem in the volatility process. The T-GARCH, first, proposed by Tong (1990) and extended by Zakoin (1994) accommodates structural breaks in volatility. It allows describing the regime shifts in the volatility, expressed as follows:

$$\sigma_t^2 = \omega + \sum_{i=1}^{q}(\alpha_i|\varepsilon_{t-i}| + \gamma_i|\varepsilon_{t-i}^+|) + \sum_{i=1}^{p}\beta_j\sigma_{t-j} \tag{23}$$

Where $\alpha_i$, $\beta_j$, $\omega$ and $\gamma$ are the parameters to estimate.

The *ETH* price volatility follows an explosive process since the duration of persistence sharply exceeds 1 ($\alpha + \beta + 0.5\gamma = 1.19$). The leverage effect is positive and statistically significant implying that the conditional variance reacts to bad news rather than good news.

In sum, we deduce that both Bitcoin and Ether are highly volatile and more responsive to bad news, but the volatility of Ether is more persistent than that of Bitcoin. As virtual currencies, both Bitcoin and Ether may be associated to multiple risks. Bitcoin, for instance, is sensitive to cyber-attacks that may play a destabilizing role in its system (Matonis 2012, Moore and Christin 2013). Bitcoin suffers also from information asymmetry, as its system is relatively complex and thus may not be easily understood by all users (Ciaian et al. 2016). The fact that Ether appears more volatile than Bitcoin may be due the acceptance and the awareness of Bitcoin compared to the nascent Ether.

**Table 3. Volatility' parameters and persistence**

|  | **BPI** E-GARCH | **ETH** T-GARCH |
|---|---|---|
| Dependent variable: ($r_t$) | | |
| Mean equation | | |
| $C$ | -0.0373*** (0.0000) | -0.0942*** (0.0000) |
| $r_{t-1}$ | -0.4214* (0.0124) | -0.4256** (0.0078) |
| Variance equation | | |
| $\omega$ | -0.0345 (0.1115) | -0.0119*** (0.0000) |
| $\alpha$ | 0.794* (0.0161) | 0.314* (0.0456) |
| $\beta$ | -0.047 (0.8741) | 0.778*** (0.0000) |
| $\gamma$ | 0.2587 (0.6130) | 0.1968 (0.4167) |
| The duration of persistence: $\alpha + \beta + 0.5\gamma$ | 0.876 | 1.190 |
| The leverage effect: $\gamma$ | 0.258 | 0.196 |

Notes: $\omega$: the reaction of conditional variance; $\alpha$: the ARCH effect; $\beta$: the GARCH effect; $\gamma$: the leverage effect; $r$: is the return of *BPI* and *ETH*.

### 3.2. Speculation

SADF and GSADF tests developed by Phillips et. al. (2011, 2013) were carried out to determine Bitcoin and Ether price bubbles, and in turn test their speculative behaviors. Monte Carlo simulation was explored with 10000 iteration while test statistics were being acquired during analysis. The initial window size was set as 0.1. The outcomes were obtained through trend and intercept models due to the structure having trend of prices. According to results of the study reported in Table 4, bubbles came onto being for Bitcoin prices. This does not hold for Ether prices. Unlike the SADF test statistic related to *ETH*, the SADF test statistic related to *BPI* were higher than critical values verifies this finding. Likewise, according to GSADF test, we notice the existence of bubbles in *BPI* because the test statistics are below the critical values.

**Table 4. Results of SADF and GSADF Test for Bitcoin Price Index and Ether price index (n=0.1)**

| | Test statistic | |
|---|---|---|
| | SADF | GSADF |
| **BPI** | 0.055* | 0.083* |
| **ETH** | 0.0062 | 0.0074 |
| Critical values | | |
| | 1% | 5% | 10% |
| SADF | 3.85 | 1.05 | 0.05 |
| GSADF | 9.75 | 4.50 | 1.95 |

Note: *, ** and ***: indicates significance at the 10%, 5% and 1% levels.

The detailed findings in Table 4 can be depicted in Figure 1 of SADF and GSADF unit root test. From Figure 1, we clearly note that *BPI* prices increase above the average successively. These sharp ups cause bubbles as they not disappear in short time-horizons. Nowadays, speculators are driving a cryptocoin bubble. Basically, from 2009 to 2017, the Bitcoin price went from a fraction of a cent, to $1000 then plunged to $300 to ultimately attained more than $2000. This was mainly attributed to the increased uncertain political atmosphere around the globe that led investors to escape to hedging tool and safe haven. Ether began at around $7 in the end of 2015 and grew progressively to a maximum price of $40 to 50 where it remained more or less stable until last month. At the start of May 2017, it skyrocketed to $170 and then to $250 as of today. This may be explained by a spread general awareness around the benefits of a next generation decentralized internet. A potential element that may explain the existence of bubbles in *BPI* is its higher dependence to media coverage. Indeed, the alteration of positive and negative news contributed to high Bitcoin price cycles. However, the speculation does not drive significantly Ether prices. This can be due to the fact that compared to Bitcoin, Ethereum (Ether) is younger and its community is smaller composed by developers rather than speculators.

**Figure 1. A detection of bubble-periods in Bitcoin and Ether price returns**

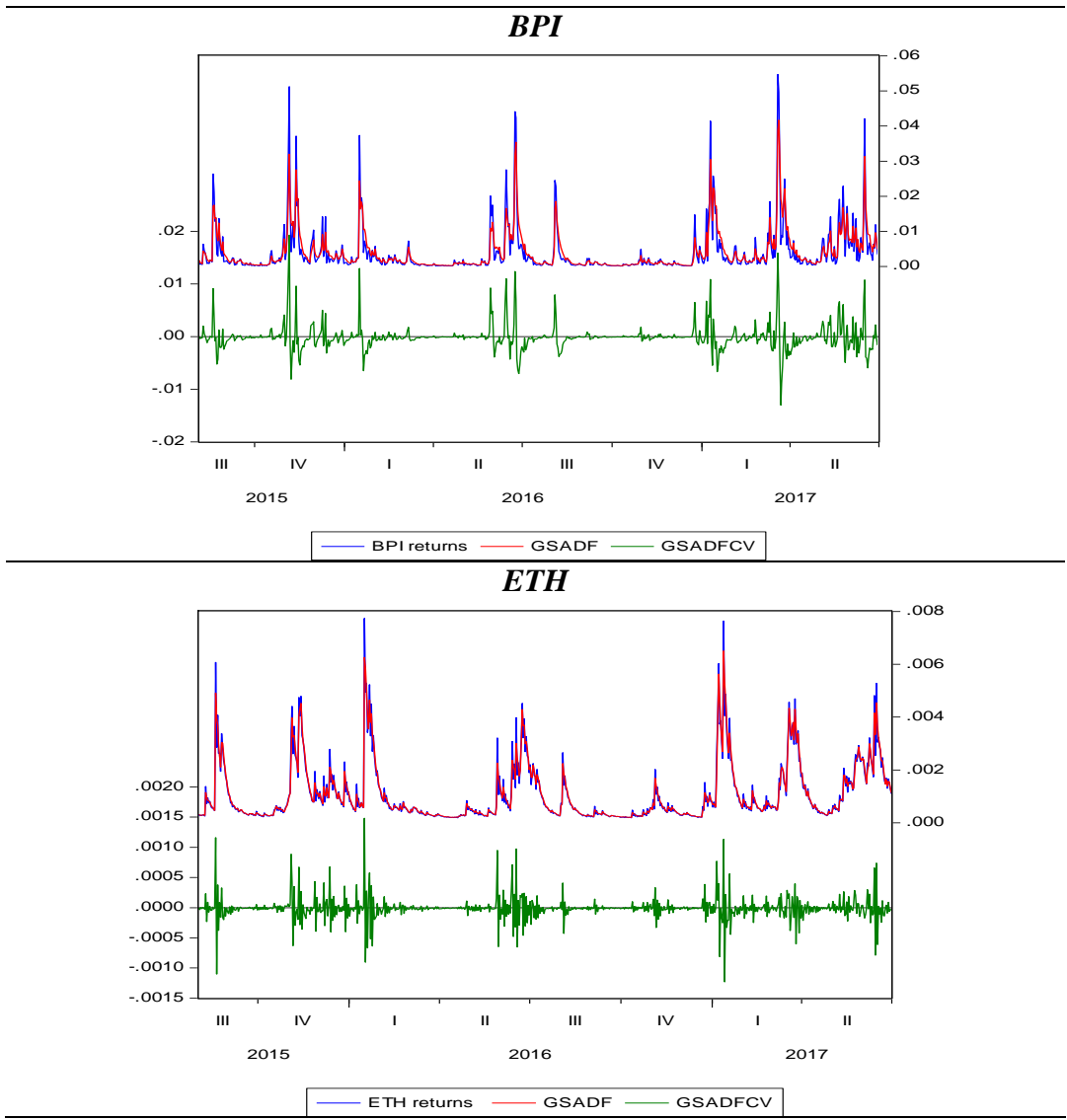

## 3.3. Hedge/Safe haven

When thinking about safe haven and hedge capabilities, investors instinctively focus on correlation. The latter provides insights about the sign and the strength of the relationship among the returns of two investments. The diversification does not ensure systematic gains, but it allows mitigating the untoward risks, leaving the investor or the trader less at the mercy of market extremes. Theoretically, a strong (weak) safe haven is defined as an asset that has a positive (negative) return in periods where another asset is in distress, while a hedge is an asset that is negatively correlated or uncorrelated with the performance of assets on average. Table 5 displays the parameter estimates for the copula models described above. The time-varying and asymmetric copula models perform better than symmetric and the time-invariant dependence copulas in most cases on the basis of the AIC outcomes and almost for all the investigated pairs. We find a strong and negative correlation between Bitcoin price and all the assets under study. Similarly for Ether price. Our findings also indicate that the lower tail dependence parameter ($\psi_1$) is stronger than the upper tail dependence parameter ($\psi_2$) for almost all the cases, which highlights that Bitcoin/Ether prices and asset (stocks, bonds and oil) prices co-move more strongly in the bearish mode rather than in the bullish state. Thus, our results document that both Bitcoin and Ether serve as a hedge, but also act as a weak safe haven.

**Table 5. Copula estimates: Correlation between digital currencies (Bitcoin vs. Ether) and assets**

|  | *BPI-STR* TV-Rotated Copula | *ETH-STR* TV-Rotated Copula | *BPI-BdR* TV-Gumbel Copula | *ETH-BdR* TV-Gaussian Copula | *BPI-Oil* TV-Rotated Copula | *ETH-Oil* TV-Gumbel Copula |
|---|---|---|---|---|---|---|
| $\psi_0$ | -0.190* (0.053) | 0.096 (0.154) | 0.345 (0.957) | 0.367 (0.234) | 1.057 (0.531) | 0.021 (0.147) |
| $\psi_1$ | -0.022** (0.008) | -0.156** (0.004) | -0.002* (0.0275) | -0.123** (0.008) | -0.045* (0.067) | -0.058* (0.034) |
| $\psi_2$ | -0.015 (0.386) | -0.081* (0.021) | -0.011* (0.052) | -0.010*** (0.000) | -0.014** (0.078) | -0.056* (0.045) |
| AIC | -35.42 | -32.82 | -32.05 | -30.96 | -36.95 | -40.07 |

Note: The table shows the maximum likelihood estimates for the different copula models for the prices of Bitcoin and Ether and S&P500 stocks, US bonds and oil price. The p-values are presented in the brackets and Akaike information criterion (AIC) values adjusted for small-sample bias are provided for the different copula models; *, ** and *** : indicates significance at the 10%, 5% and 1% levels.

### 3.4. Risk diversification results

Another potential reason behind investing in digital currencies (in particular, Bitcoin) is portfolio diversification because Bitcoin is proven by several studies to be a good diversifier (for example, Dyhrberg 2016; Bouri et al. 2017). What does this mean? In general, the diversification helps investors to achieve a desired return without highly risking as with an individual security. We try in the following to test whether Bitcoin and Ether are good diversifiers, and in turn help to mitigate the risk in a portfolio. Table 6 reports the risk evaluation findings at the 99% confidence level using the best fitting copula each pair. The risk reduction results indicate that the weighted portfolio 2 (*BPI-STR*) stocks and portfolio 3 (*ETH- STR*) can help investors lighten risk much more than the portfolio 1 composed only by S&P 500 stocks. Likewise, by introducing Bitcoin and Ether in the US bonds portfolios, market participants reach a sharp risk reduction, i.e., portfolio 5 (*BPI-BdR*) and portfolio 6 (*ETH-BdR*) versus portfolio 4 composed only be *BdR*. This holds also true when including digital currencies in a portfolio composed only by oil, i.e., portfolio 7 versus portfolio 8 (*BPI-Oil*) and portfolio 9 (*ETH-Oil*). These findings underscore that including virtual currencies in a portfolio may enhance its efficiency. This holds true for different assets: stocks, bonds, oil. Furthermore, the conditional coverage test implies that the portfolio composed of Bitcoin and *STR*/*BdR* perform better than that formed by Ethereum and *STR*/*BdR* since the null of correct conditional coverage cannot be rejected at the 5% significance level. This implies that investors who hold diversified portfolios made of stocks or bonds and Ether may face substantial losses over bear state, due to the stronger dependence in the lower tails of their return distributions. In such situation, stock and bond investors may take a short position on Bitcoin (much less pronounced dependence between *BPI* and *STR/ BdR,* see $\psi_1$ in Table 3) in order to avoid extreme losses.

**Table 6. Risk evaluation for digital currencies (Bitcoin vs. Ether) and assets**

|  | Portfolio 2 vs. Portfolio 1 | Portfolio 3 vs. Portfolio 1 | Portfolio 5 vs. Portfolio 4 | Portfolio 6 vs. Portfolio 4 | Portfolio 8 vs. Portfolio 7 | Portfolio 9 vs. Portfolio 7 |
|---|---|---|---|---|---|---|
| Risk Red. | 0.5311*** | 0.4921*** | 0.1805** | 0.1978* | 0.2910* | 0.4516** |
| Cond. Cov. | 0.4762** | 0.0876 | 0.3134** | 0.1511 | 0.1689* | 0.3194*** |

Notes: This table display the risk evaluation outcomes for portfolios composed of digtal currencies and assets (stocks, bonds, oil) compared to benchmark portfolios composed only of stocks (or only of bonds or only for oil). Risk Red indicates the risk effectiveness ratio; Cond cov indicates the *P* values for the conditional coverage test. *, ** and ***: indicates significance at the 10%, 5% and 1% levels.

## 4. Concluding remarks

Globalization has led many to take a particular attention in financial services that are agnostic to national borders, including cryptocurrencies. The number of people who utilize virtual currencies is steadily rising, and excitement has increased around the possibility that the price of Bitcoin and Ether will surge remarkably in coming years. Indeed, the conditions are in place to make Bitcoin and Ether poised to play a potential role in the World economy[6]. Nowadays, Bitcoin and Ether are both experiencing a boom, but are they complementary or competitors with each other? Can it objectively be asserted one is better than the other, or do they have different properties? Some compare Bitcoin and Ether to "lions" on land and "sharks" in the water which are at the top of their respective food chains, but not in competition with each other. But also several cryptocurrency analysts expect that ether's value could overtake Bitcoin price by the end of 2018. According to Coinmarketcap data, Bitcoin dominates now less than half of the digital currency market, and Ether represents about a quarter of the total cryptocurrency market. Ethereum's supporters think that Ethereum could become a globally accessible vehicle for running businesses as the technology enables more intricate transactions in a decentralized way. This study goes beyond a simple comparison and seeks to better understand the differences between these two cryptocurrencies. Using several methods (Optimal GARCH model, the generalized sup ADF test procedure, copulas and Reboredo (2013)'s measures), we compare the Bitcoin and Ether across their volatile attitudes, their speculative behaviors, and other properties (i.e., acting as safe haven, hedge and risk diversifier). Quite interesting findings were drawn.

---
[6] For example, the mobile device adoption which eases digital currency transfer continues to rise across the globe.

We document that both Bitcoin and Ether are highly volatile, but Bitcoin seems *less volatile* and *more speculative* than Ether. A challenge for most digital currency owners is that they do not have a background in traditional investing, and, therefore do not have all wisdom on how to effectively handle volatility. The point is that the speed of change in cryptocurrency markets is simply much times higher than other markets. Because of the infancy of Ethereum's platform, Ether may be linked to more risks than Bitcoin. Another element that may explain why Ether is more volatile than Bitcoin is that Ethereum's blockchain enables blocks to be mined extremely quickly with a block time of 14 seconds, compared to Bitcoin's block time of 10 minutes, which ensures stronger transactional velocity. Moreover, the capping of Bitcoin's supply at 21 million brings a controlled supply which improves its predictability compared to Ether, which has no hard limit. Add to this that Bitcoin has the advantage with respect its wider acceptance and awareness and its infrastructural presence, yielding to less pronounced volatility. Interestingly, after eight years of development, the Bitcoin network is always depicted by supporters as the most secure blockchain. Regardless of the positive views regarding its powerful network effects and diverse mining network, Ethereum has faced great criticism with respect security problems for various reasons, especially the fact that the software is in its nascent stages and has only been available for two years.

Our findings also indicate that there Bitcoin/Ether prices are negatively correlated with financial assets (in particular, S&P500 stocks, US bonds, oil), which underscore that cryptocurrencies can act as *hedge and weak safe haven* against the price movements of these assets. Investors and traders are generally interested in hedges that mitigate the volatility of their portfolio, but also they are likely interested in buying some sort of insurance against extreme tail events. Digital currencies have several properties that make them useful for both cases. Currently, Bitcoin and Ether –which live outside the confines of a single country's politics– profited from the great uncertainty heightened across the globe and the loss of faith in the stability of banking system. Our results indicate also that digital currencies are unlikely to act as strong safe haven. In general, investors tend to sell "risky" assets and buy "safer" assets in periods of great uncertainty (Baele et al. 2015). Both Bitcoin and Ether, however, do not hold this characteristic. From a legal perspective, Bitcoin and Ether do not appear to share the features of traditional safe-haven investments (gold, for instance). The ability of Bitcoin and Ether to act as safe haven may encounter a number of obstacles, especially with regard their volatile and speculative behaviors demonstrated above.

Furthermore, our results suggest that the inclusion of Bitcoin or Ether into diversified portfolio may be profitable, *serving therefore as risk diversifiers*. Nevertheless, the investors who hold portfolios containing stocks/bonds and Ether may face great losses during bear states. In such context, stock and bond investors may turn to Bitcoin. But prior to making such investment, we shouldn't forget to mention the major challenges facing investors in digital monies. Given the short track record of these assets, there is not a standard valuation tool that is largely accepted to predict the trading prices of Bitcoin and Ether, and there is no consensus on the best method able to estimate the price trend (Gosh et al. 2017). Moreover, the cryptocurrency market is exposed to serious speculations, and new players enter the market every day, making the application of any valuation method problematic.

Last but not least, several cryptocurrencies provide an alternative to Bitcoin without offering any clear reason to switch. Ether is the only alternative that comes with a various set of advantages especially because of Ethereum's smart contracts. While Bitcoin and Ether are meant for distinct purposes[7] and co-exist in the industry, they do compete with each other for getting the maximum number of users. If Ethereum's blockchain succeeds to effectively bring out a revolution in the adoption of smart contracts, interest is expected to grow, and the adoption will spread, and as more people use Ethereum, Ether's prices should rise.

---

[7] While Bitcoin is developed as an alternative to regular money and is thus a medium of payment transaction and store of value, Ethereum is created as a platform which eases peer-to-peer contracts and applications via its own currency vehicle (Ether). Both Bitcoin and Ether are based on blockchains, but Ethereum's blockchain extends the concept of a distributed ledger to enable further advanced commands.

# Appendix

## Table A.1. Summary of the differences between Bitcoin and Ethereum

|  | **Bitcoin** | **Ethereum** |
|---|---|---|
| Purpose | Bitcoin is a currency | Ethereum is a platform for running decentralized applications (i.e., smart contracts) |
| Supply | The Bitcoin is deflationary : the Bitcoin supply is limited | The Ethereum is inflationary: The Ethereum supply in unlimited |
| Security | The built-in language is not Turing-complete, implying that there are some programs that are impossible to write. | Ethereum has a rich programming language built-in. The built-in language is Turing-complete, implying that you can code anything you want. |
| Mainstream adoption | The community is bigger. | Ethereum is less known and younger (the Ethereum community is still not larger). |

## Table A.2. GARCH models used in this study

| |
|---|
| GARCH-M (GARCH in mean, Bollerslev et al. 1993) $$r_t = \mu_t + \varepsilon_t + \lambda \sigma_t^2$$ |
| C-GARCH (Component GARCH, Ding et al. 1993) $$(\sigma_t^2 - \sigma^2) = \alpha(\varepsilon_{t-1}^2 - \sigma^2) + \beta(\sigma_{t-1}^2 - \sigma^2)$$ |
| I-GARCH (Integrated GARCH, Bollerslev et al. 1993) $$\sigma_t^2 = \omega + \varepsilon_{t-1}^2 + \sum_{i=1}^{q} \alpha_i (\varepsilon_{t-i}^2 - \varepsilon_{t-1}^2) + \sum_{j=1}^{p} \beta_j (\sigma_{t-j}^2 - \varepsilon_{t-1}^2)$$ |
| T-GARCH (Threshold GARCH, Zakoian, 1994) $$\sigma_t^2 = \omega + \sum_{i=1}^{q} (\alpha_i |\varepsilon_{t-i}| + \gamma_i |\varepsilon_{t-i}^+|) + \sum_{j=1}^{p} \beta_j \sigma_{t-j}$$ |
| E-GARCH (Exponential GARCH, Nelson, 1991) $$\log(\sigma_t^2) = \omega + \sum_{i=1}^{q} (\alpha_i z_{t-i} + \gamma_i (|z_{t-i}| - \sqrt{2/\pi})) + \sum_{j=1}^{p} \beta_j \log(\sigma_{t-j}^2)$$ |
| P-GARCH (Power GARCH, Higgins and Bera, 1992) $$\sigma_t^\varphi = \omega + \sum_{i=1}^{q} \alpha_i \varepsilon_{t-i}^\varphi + \sum_{j=1}^{p} \beta_j \sigma_{t-j}^\varphi$$ |
| A-PGARCH (Asymmetric power GARCH, Ding et al., 1993) $$\sigma_t^\varphi = \omega + \sum_{i=1}^{q} \alpha_i (|\varepsilon_{t-i}| + \gamma_i \varepsilon_{t-i})^\varphi + \sum_{j=1}^{p} \beta_j \sigma_{t-j}^\varphi$$ |
| CMT-GARCH (Component with Multiple Thresholds GARCH, Bouoiyour and Selmi, 2014) $$\sigma_t^2 = \omega + \alpha \varepsilon_{t-1}^2 + \beta(\omega + (\alpha + \gamma I_{(\varepsilon_{t-2}<0)})\varepsilon_{t-2}^2 + \beta \sigma_{t-2}^2)$$ |

Notes: $\sigma_t^2$: conditional variance, $\alpha_0$: reaction of shock, $\alpha_1$: ARCH term, $\beta_1$: GARCH term, $\varepsilon$: error term; $I_t$: denotes the information set available at time t; $z_t$: the standardized value of error term where $z_t = \varepsilon_{t-1}/\sigma_{t-1}$; $\mu$: innovation, $\gamma$: leverage effect; $\varphi$: power parameter.

**Table A.3. Optimal GARCH models chosen via information criteria**

| Models | BPI | ETH |
|---|---|---|
| | Akaike criterion | |
| GARCH | -5.7652 | -5.9873 |
| GARCH-M | -5.6891 | -5.6233 |
| I-GARCH | -5.1683 | -5.2345 |
| C-GARCH | -5.6428 | -4.8972 |
| CMT-GARCH | -5.6442 | -5.0761 |
| T-GARCH | -5.8136 | -5.9914 |
| E-GARCH | -5.9364 | -5.6370 |
| P-GARCH | -5.7812 | -5.5043 |
| AP-GARCH | -5.7261 | -5.7261 |
| | Bayesian criterion | |
| GARCH | -5.4463 | -5.5231 |
| GARCH-M | -5.4374 | -5.5295 |
| I-GARCH | -5.4453 | -5.5367 |
| C-GARCH | -5.3076 | -5.4151 |
| CMT-GARCH | -5.6731 | -5.7071 |
| T-GARCH | -5.6021 | -5.9942 |
| E-GARCH | -5.7376 | -5.5669 |
| P-GARCH | -5.4639 | -5.5406 |
| AP-GARCH | -5.1067 | -5.2295 |
| | Hannan-Quinn criterion | |
| GARCH | -5.3957 | -5.4234 |
| GARCH-M | -5.3682 | -5.4015 |
| I-GARCH | -5.5387 | -5.5553 |
| C-GARCH | -5.3512 | -5.3923 |
| CMT-GARCH | -5.3934 | -5.4263 |
| T-GARCH | -5.4131 | -5.7524 |
| E-GARCH | -5.6521 | -5.3966 |
| P-GARCH | -5.4017 | -5.4294 |
| AP-GARCH | -5.3816 | -5.4149 |

Notes: The model with the minimum value is assumed to be the optimal one. The formula of the different historical evaluation used in this study can be written as follows: Akaike information criterion : *-2log (vraisemblance) + 2k*; Bayesian information criterion : *-2log (vraisemblance) + log(N).k*; Hannan-Quinn information criterion: *-2log (vraisemblance) + 2k.log (log(N))* where *k* the degree of freedom and *N* is the number of observations.